\documentclass[11pt]{article}
\usepackage[margin=2.5cm]{geometry}
\usepackage{amsmath,amssymb,amsthm,mathtools}

\usepackage[draft]{fixme}
\fxsetup{theme=color, mode=multiuser, layout=inline}
\FXRegisterAuthor{ms}{aMS}{Maciej}
\FXRegisterAuthor{og}{aOG}{Onur}

\usepackage{booktabs}
\usepackage{graphicx}
\usepackage{xcolor}

\usepackage{hyperref}
\usepackage[nameinlink]{cleveref}
\crefname{theorem}{Theorem}{Theorems}
\crefname{lemma}{Lemma}{Lemmas}
\crefname{corollary}{Corollary}{Corollaries}
\crefname{section}{Section}{Sections}

\definecolor{linkblue}{RGB}{0, 101, 189}     % Deep blue for internal links/citations
\definecolor{urlteal}{RGB}{0, 119, 139}      % Teal for external URLs
\definecolor{filegreen}{RGB}{19, 114, 70}    % Forest green for files/menus

\hypersetup{
    colorlinks=true,       % False: boxed links; True: colored text
    linkcolor=linkblue,    % Color of internal links (sections, equations)
    citecolor=linkblue,    % Color of citations
    urlcolor=urlteal,      % Color of external URLs
    filecolor=filegreen    % Color of file links
}

\definecolor{keyblue}{RGB}{242,248,253}
\newcommand{\keybox}[1]{%
  \begingroup
  \setlength{\fboxsep}{9pt}%
  \colorbox{keyblue}{\makebox[\dimexpr\linewidth-2\fboxsep\relax]{\(\displaystyle #1\)}}%
  \endgroup
}

\newtheorem{theorem}{Theorem}
\newtheorem{lemma}[theorem]{Lemma}

\newtheorem{corollary}[theorem]{Corollary}
\theoremstyle{remark}
\newtheorem{remark}[theorem]{Remark}

\newcommand{\E}{\mathbb{E}}

\title{Birthday Estimators Are Uniformly Optimal for R\'enyi Entropy}
\author{Maciej Sk\'orski}
\date{\today}

\begin{document}
\maketitle

\begin{abstract}
Collisions are a basic probe of unknown distributions, used for testing and for counting substructures in graphs, and they underlie R\'enyi entropy, a fundamental measure of uncertainty with applications in cryptography, machine learning, physics, and biology. 
Motivated by this, we study the natural birthday estimator. 
We prove Bernstein-type concentration for it and show that it is minimax optimal, uniformly and jointly in alphabet size, order, accuracy, and confidence, with no sample splitting, clipping, or median aggregation.

This, in particular, settles the sample complexity of integer-order R\'enyi entropy estimation up to absolute constants in all regimes, closing the gap between the best known upper and lower bounds. 
Our concentration bound rests on a sensitivity analysis of collision count, and the matching lower bound on a sharp analysis of detecting a hidden spike in a plateau distribution.
\end{abstract}

\section{Introduction}

\subsection{Background}

% collisions are important...
Sample collisions are a basic tool for probing an unknown distribution. 
They have numerous applications, such as uniformity and closeness testing~\cite{GR00,DGPP19}, security of hash functions~\cite{STKT06}, counting substructures in graphs and hypergraphs~\cite{JSP13,XB24}, and most notably determining R\'enyi entropy, a fundamental measure of uncertainty of central importance for cryptography~\cite{BBCM95,HILL99}, machine learning~\cite{Prin10}, physics~\cite{BEL19,BS93}, and biology, where it quantifies diversity~\cite{Hill73,Jost06}. 

% .. estimation seems easy by birthdays!
The order-$d$ collision probability is the chance that $d$ independent samples coincide,
$$
C_d(p)=\Pr_{x_1,\ldots,x_d\overset{}{\sim}p}(x_1=\cdots=x_d).
%NOTE: no need for underscoring i.i.d. 
$$
Given this operational definition, a \emph{birthday estimator} naturally comes to mind: \emph{just count coincidences} in the sample, and normalize by the number of $d$-subsets. 
Since R\'enyi entropy $H_d$ is a logarithmic rescaling of $C_d$, the same estimator applies~\cite{AOST17,OS17,Sko23}.

% problem: birthday estimators easy to do hard to analyze!
Yet the uncertainty of this simple method is not fully understood: coincidences overlap heavily, so that many dependent events are counted.
To guard against potentially heavy tails, existing general-order analyses resort to independent batches and median aggregation~\cite{OS17,Sko23}, sometimes with clipping~\cite{AK26}, rather than directly controlling the estimator. 
Because of these difficulties, the optimal sample size for collision and entropy estimation is not fully settled: the joint dependence on order, accuracy, confidence, and alphabet size remains open~\cite{AK26}.
% no need to credit ~\cite{AOST17} here 
This work revisits this problem:
\begin{quote}
\emph{Does the birthday estimator concentrate sharply enough to yield optimal entropy estimation, jointly in order, accuracy, confidence, and alphabet size?}
\end{quote}

\subsection{Contributions.}
We develop strong concentration bounds for the birthday estimator, and conclude that it gives high-confidence R\'enyi entropy estimation that is optimal jointly in alphabet size, integer order, accuracy, and confidence. 
Throughout, $p$ is a distribution on $k$ symbols, $d\ge2$ is an integer order, $\varepsilon$ the accuracy, $\delta$ the failure probability, and $K=k^{1-1/d}$; the symbols $\lesssim,\asymp$ hide absolute constants.

\paragraph{Strong concentration of birthday statistics.}
Collision counts have self-bounding properties. Through a sensitivity analysis of the suitably normalized and regularized count, we show that the number $\widehat C_d$ of $d$-way coincidences among $n$ samples satisfies, for every distribution, integer $d\ge2$ and $t>0$, with probability at least $1-2e^{-t}$,
$$
\keybox{
\big|\widehat C_d^{\,1/d}-m\big|\lesssim 1+\sqrt{\frac{mt}{d}}+\frac td,
\qquad m=(\E\,\widehat C_d)^{1/d},
}
$$
a Bernstein-like bound around the root $m$ of its mean, with variance proxy of order $m/d$ and scale $1/d$, obtained without sample splitting, clipping, or median aggregation (\Cref{thm:concentration}).

\paragraph{Learning R\'enyi entropy.}
Transferring this bound to the logarithmic scale, the unchanged order-$d$ estimator learns $H_d(p)$ to additive accuracy $\varepsilon$ with probability at least $1-\delta$ from
$$
\keybox{
n\gtrsim
\underbrace{\|p\|_d^{-1}}_{\text{distribution scale}}
\left(
\frac d\varepsilon+\frac{\log(1/\delta)}{\varepsilon^2}
\right)
}
$$
samples (\Cref{thm:estimator}), where $\|p\|_d^{-1}=C_d(p)^{-1/d}\le K$. This guarantee is distribution-dependent.

\paragraph{Learning hidden spikes.}
Of independent interest, we study a plateau of $r$ atoms of mass $1/K$ in which one randomly chosen atom is raised by a relative amount $\eta$, the excess being compensated by a uniform remainder. Distinguishing this from the plain plateau with error at most $\delta$ requires, and for $\eta\le1$ suffices,
$$
\keybox{
n\asymp
\underbrace{K\eta^{-2}}_{\text{relative-change detection}}
\left(
\underbrace{\log(1/\delta)}_{\text{confidence}}
+
\underbrace{\log r}_{\text{search cost}}
\right)
}
$$
samples: the lower bound follows from an exact finite-sample mixture $\chi^2$ calculation, without Poissonization, and the upper bound from a simple max-count test (\Cref{thm:plateau_testing}).

\paragraph{Universal minimax optimality.}
Choosing $\eta\asymp\varepsilon$ and $\log r\approx\min\{d\varepsilon,\log K\}$ turns the spike bound into an entropy lower bound that matches the upper bound. The minimax sample complexity $n^*(k,d,\varepsilon,\delta)$ of additive R\'enyi entropy estimation, the least $n$ for which some estimator is $\varepsilon$-accurate with probability $1-\delta$ on every distribution over $k$ symbols, therefore satisfies
$$
\keybox{
n^*(k,d,\varepsilon,\delta)\asymp
\frac{K}{\varepsilon^2}
\left(
\underbrace{\min\{d\varepsilon,\log K\}}_{\text{search cost}}
+
\underbrace{\log(1/\delta)}_{\text{confidence}}
\right),
}
$$
uniformly in $k\ge16$, $d\ge2$, and $\varepsilon,\delta$ below small absolute constants (the nontrivial regime), with the order capped when necessary (\Cref{thm:main}).

\subsection{Related Work}

\paragraph{Sample complexity of R\'enyi entropy.}
Acharya, Orlitsky, Suresh and Tyagi~\cite{AOST17} showed that integer orders $d>1$ need only $n\asymp_d k^{1-1/d}$ samples (constants depending on $d$), against near-linear in $k$ for non-integer orders above one and $k/\log k$ for Shannon entropy~\cite{VV11,JVHW15,WY16}.
The follow-up work~\cite{OS17} replaced the alphabet size by the scale $\|p\|_d^{-1}$, which we inherit; the analysis was later simplified with the theory of U-statistics and made adaptive~\cite{Sko23}; a relative error $\varepsilon$ for $P_d$ costs $\|p\|_d^{-1}d^2\varepsilon^{-2}\log(1/\delta)$ samples, with confidence boosted by a median of blocks. 
For entropy accuracy $\varepsilon$ this is $\|p\|_d^{-1}\max\{d^2,\varepsilon^{-2}\}\log(1/\delta)$: optimal for $d\varepsilon\lesssim1$ but lossy beyond.
Adibi and Krysta~\cite{AK26} prove $Kd\varepsilon^{-2}\log(1/\delta)$; this work replaces this product by the sum $K(d/\varepsilon+\varepsilon^{-2}\log(1/\delta))$ and matches it in all regimes from below.

\paragraph{Direct concentration inequalities.}
A standard route to high-confidence guarantees for $U$-statistics combines a variance bound with median aggregation~\cite{JVV86}, multiplying the sample size by $\log(1/\delta)$; another uses Hoeffding's blocking argument~\cite{Hoeffding63}, with effective sample size $\lfloor n/d\rfloor$. 
% blocking not good
For collision counts, blocking combined with Bernstein gives $n\gtrsim dP_d^{-1}\varepsilon^{-2}\log(1/\delta)$ for relative error $\varepsilon\in(0,1]$, whereas our analysis gives $n\gtrsim d^2P_d^{-1/d}\varepsilon^{-2}\log(1/\delta)$: for uniform $p$, the alphabet factor drops from $k^{d-1}$ to $k^{1-1/d}$.
% collisions not good either: too broad around the mean and bad - super-linear - on confidence!
General $U$-statistic inequalities do not directly supply our guarantees. For order two, \cite{GLZ00} gives a multi-regime inequality with a quadratic log-confidence term; for higher orders, its moment bounds for nonnegative kernels give only constant-factor control of the $d$-th root of the collision count, which is insufficient for small relative error, and its centered versions carry order-dependent constants that are not made explicit.
%NOTE: prefer not cite myself in here. For pair collisions ($d=2$), direct bounds without boosting were shown in~\cite{Sko21}; for general $d$ none seems known. 
We obtain direct high-confidence bounds for every integer $d$ by self-bounding concentration~\cite{BLM00,BLM09,Maurer06} of the collision count (normalized and regularized), with variance proxy $\lesssim\mu/d$ and scale $\lesssim1/d$, and absolute constants (\Cref{thm:concentration}).

\paragraph{Lower bounds.}
Hidden-coordinate mixtures were used in~\cite{AK26} to obtain hard instances for entropy estimation, with Poissonization and truncation in the analysis.
We use a different three-level family (floor, plateau, hidden spike) and compute the $\chi^2$-divergence of the mixture exactly at fixed $n$; a matching test shows that the resulting bound is tight up to a constant (\Cref{thm:plateau_testing}).

\section{Preliminaries}

We set up the notation and precisely define the notions below.

\subsection{Collision probability and R\'enyi entropy}
Write $[k]=\{1,\dots,k\}$ and let $\mathcal P_k$ be the distributions on $[k]$. For $p\in\mathcal P_k$ and an integer $d\ge2$,
$$
P_d(p)=C_d(p)=\sum_xp_x^d,\qquad \|p\|_d=P_d^{1/d},\qquad H_d(p)=\frac{-\log P_d(p)}{d-1}=\frac d{d-1}\log\frac1{\|p\|_d},
$$
and $K=k^{1-1/d}$. Logarithms are natural. The entropy $H_d$ is nonincreasing in $d$, with $H_\infty=-\log p_{\max}\le\log k$. By the power-mean inequality $P_d\ge k^{1-d}$, so the distribution scale $\|p\|_d^{-1}$ never exceeds $K$, with equality for the uniform distribution.

\subsection{The birthday estimator}
$X_1,\dots,X_n$ are i.i.d.\ from $p$, with $n\ge d$ and counts $N_x$, and $\Pr,\E$ refer to this sample. The sample collision count is
$$
\widehat C_d=\sum_{1\le i_1<\cdots<i_d\le n}\mathbf1\{X_{i_1}=\cdots=X_{i_d}\}
=\sum_x\binom{N_x}{d},\qquad \binom Nd=0\ \text{for}\ N<d.
$$
Each indicator has expectation $C_d(p)=P_d(p)$, hence $\E\widehat C_d=\binom nd P_d$ and $\widehat C_d/\binom nd$ is unbiased for $P_d$.
The birthday entropy estimator is
$$
\widehat H_d=\frac1{d-1}\log\frac{\binom nd}{\widehat C_d},
$$
with $\widehat H_d=+\infty$ when $\widehat C_d=0$. 

\subsection{Minimax conventions}
$n^*(k,d,\varepsilon,\delta)$ is the least $n$ for which some estimator is $\varepsilon$-accurate with probability $1-\delta$ for every $p\in\mathcal P_k$. We write $a\lesssim b$ if $a\le Cb$, and $a\asymp b$ if both $a\lesssim b$ and $b\lesssim a$, for an absolute constant $C$.

\subsection{Elementary inequalities}
For integers $N\ge d\ge1$ we use $\binom Nd\ge(N/d)^d$, the identities $\binom{N-1}{d-1}=\frac dN\binom Nd$ and $\binom Nd-\binom{N-1}{d-1}=\binom{N-1}{d}$, the bound $\binom{N-1}{d-1}\le N^{d-1}/(d-1)!$, and $d!\ge(d/e)^d$.

\section{Results}

%NOTE: results only, proofs after in their own section
We first state the minimax rate, then the concentration and estimation results behind the upper bound, and finally the hidden-spike testing result behind the lower bound.

\begin{theorem}[Minimax sample complexity]\label{thm:main}
There are absolute constants $c,C>0$ such that for all $k\ge16$, integers $d\ge2$, $0<\varepsilon\le 1/64$ and $0<\delta\le 1/8$, with $K=k^{1-1/d}$,
\[
c\,K\Big(\frac{\min(d,\ \varepsilon^{-1}\log K)}{\varepsilon} + \frac{\log(1/\delta)}{\varepsilon^2}\Big)
\;\le\; n^*(k,d,\varepsilon,\delta) \;\le\;
C\,K\Big(\frac{\min(d,\ \varepsilon^{-1}\log K)}{\varepsilon} + \frac{\log(1/\delta)}{\varepsilon^2}\Big).
\]
The upper bound is attained by the birthday estimator at order $d'=\min(d,\lceil 1+2\log k/\varepsilon\rceil)$.
\end{theorem}

\paragraph{Three regimes.}
The rate in \Cref{thm:main} has three regimes, depending on the term dominating in $\frac{K}{\varepsilon^2}\big(\min\{d\varepsilon,\log K\}+\log(1/\delta)\big)$;
each is realized by a hard instance from \Cref{thm:plateau_testing}. When $d\varepsilon\lesssim\log(1/\delta)$, confidence dominates and the rate is $K\log(1/\delta)/\varepsilon^2$; the hard instance is a single heavy atom ($r\asymp1$). When $\log(1/\delta)\lesssim d\varepsilon\lesssim\log K$, the order matters: the rate is $Kd/\varepsilon$, and the hard instance hides one raised atom among $r\approx e^{d\varepsilon}$ candidates. When $d\varepsilon\gtrsim\log K$, the search cost saturates at $\log K$ and the rate is $K(\log K+\log(1/\delta))/\varepsilon^2$, with $r\asymp K$ candidates.

The upper bound relies on a novel concentration bound for the collision count, and its consequence for entropy.

\begin{theorem}[Concentration of collision counts]\label{thm:concentration}
For every distribution $p$, integers $d\ge2$ and $n\ge1$, and every $t>0$, with $m=(\E\widehat C_d)^{1/d}$,
$$
\Pr\!\left(\big|\widehat C_d^{\,1/d}-m\big|>4+2e+\sqrt{\frac{8e(m+4)t}{d}}+\frac{2et}{d}\right)\le2e^{-t}.
$$
In particular, with probability at least $1-2e^{-t}$, $\big|\widehat C_d^{\,1/d}-m\big|\lesssim1+\sqrt{mt/d}+t/d$.
\end{theorem}

\begin{theorem}[Birthday estimator]\label{thm:estimator}
For every distribution $p$, integer $d\ge2$,
$\varepsilon\in(0,1]$ and $\delta\in(0,1/2]$, a sample of size
$$
n\ \gtrsim\ \frac{1}{\|p\|_d} \Big(\frac d\varepsilon+\frac{\log(1/\delta)}{\varepsilon^2}\Big),
$$
guarantees that $\Pr\big(|\hat H_d-H_d(p)|\le\varepsilon\big)\ge1-\delta$, where $\hat H_d$ is the birthday estimator.
\end{theorem}

\begin{corollary}[Sample complexity]\label{cor:capped}
Let $p$ be supported on $k\ge2$ symbols, let $d\ge2$ be an integer, $0<\varepsilon\le1$ and $0<\delta\le1/2$, and put $K=k^{1-1/d}$. Use the capped order
$d'=\min\{d,\lceil1+2\log k/\varepsilon\rceil\}$. Then $\hat H_{d'}$ satisfies $|\hat H_{d'}-H_d(p)|\le\varepsilon$ with probability at least $1-\delta$ whenever
$$
n\ \gtrsim K\Big(\frac{\min\{d,\ \varepsilon^{-1}\log K\}}{\varepsilon}+\frac{\log(1/\delta)}{\varepsilon^2}\Big).
$$
\end{corollary}

\begin{figure}[!h]
\centering
\includegraphics[width=0.99\linewidth]{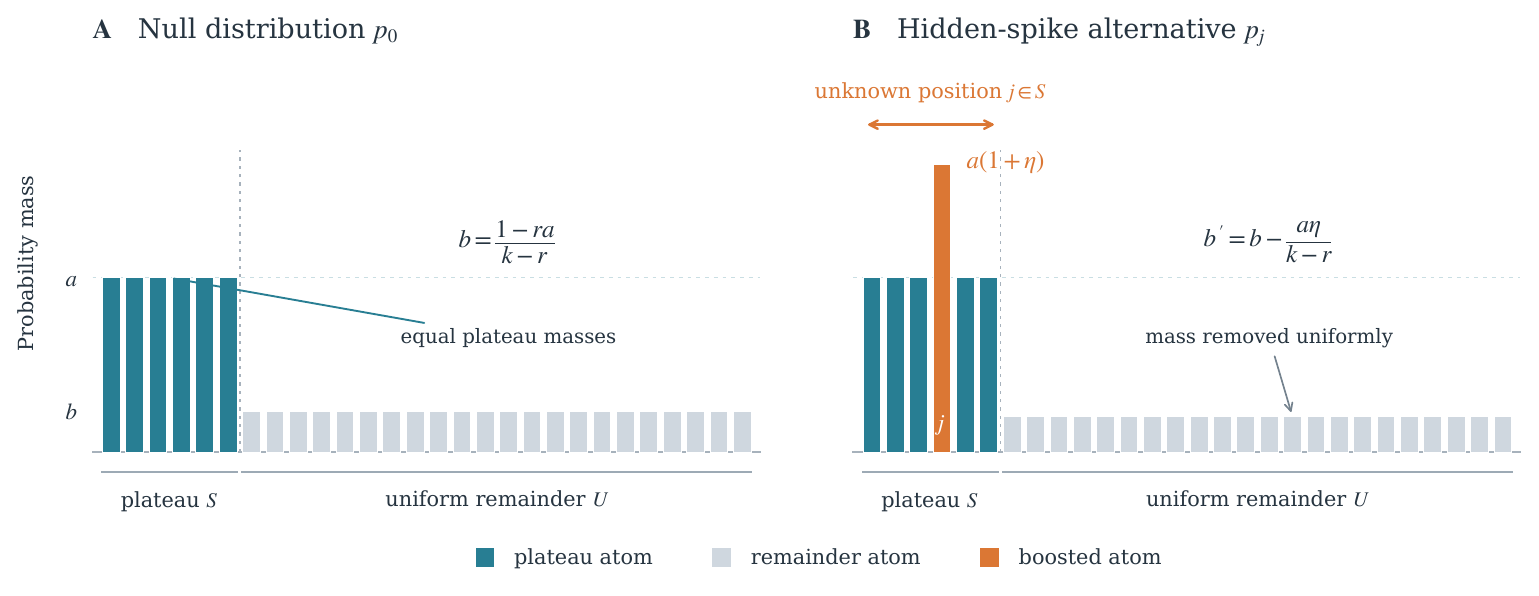}
\caption{The ``plateau'' null $p_0$ and the ``spiked'' alternative $p_j$: one random plateau atom gains mass; 
both the gain magnitude and its unknown location add to the testing difficulty.
}
\label{fig:hard-family}
\end{figure}

The lower bound will follow from the auxiliary testing problem illustrated in~\Cref{fig:hard-family}. 
It is formulated with $K$ as a free parameter, the entropy application specializes it to $K=k^{1-1/d}$.

\begin{theorem}[Distinguishing a hidden spike]\label{thm:plateau_testing}
Let $K\ge4$, and let $S\subset[k]$ be a plateau of size $1\le r\le K/2$ with $r<k$; call $[k]\setminus S$ the floor. Let $0<\eta\le K-r$ (see \Cref{fig:hard-family}).
\begin{itemize}
\item \emph{Null (plateau).} The distribution $p_0$ puts mass $1/K$ on each atom of $S$ and spreads the remaining mass uniformly over the floor. The sample law is $P=p_0^{\otimes n}$.
\item \emph{Alternative (plateau with a hidden spike).} For $j\in S$, $p_j$ raises the mass of atom $j$ by the factor $1+\eta$, lowering the mass uniformly over the floor in compensation. The spike location $j$ is sampled uniformly from $S$ and is unknown, so the sample law is $Q=\frac1r\sum_{j\in S}p_j^{\otimes n}$.
\end{itemize}
Let $0<\delta\le1/8$. Every test $\phi$ with $P(\phi=1)\le\delta$ and $Q(\phi=0)\le\delta$ requires
$$
n\gtrsim\frac{K}{\eta^2}\left(\log r+\log\frac1\delta\right).
$$
Conversely, let $\eta\le1$ and let $N_j$ be the number of samples equal to $j$. The test $\phi=\mathbf1\{\max_{j\in S}N_j\ge n(1+\eta/2)/K\}$ has both errors at most $\delta$ once
$$
n\gtrsim\frac{K}{\eta^2}\left(\log r+\log\frac1\delta\right).
$$
Hence for $\eta\le1$ the least sample size of a $\delta$-test is $\asymp\frac{K}{\eta^2}\left(\log r+\log\frac1\delta\right)$.
\end{theorem}

\section{Proofs}

\subsection{Birthday Estimation: High Confidence by Self-Bounding}\label{sec:upper}
We first control collisions on their natural root scale, and then pass to
R\'enyi entropy. The observations are independent; neither the symbol counts
nor the collision indicators need to be independent.

\paragraph{A moment bound.}
\begin{lemma}[Sub-gamma moment bound]\label{thm:subg_1side_moment}
Let \(W\ge0\) with \(\mu=\mathbb EW\) and \(\|W\|_q=(\mathbb EW^q)^{1/q}\), and let $v,c>0$. Suppose
\[
\log\mathbb Ee^{\lambda(W-\mu)}
\le\frac{v\lambda^2}{2(1-c\lambda)},
\qquad 0\le\lambda<1/c.
\]
Then, for every \(q\ge1\),
$$
\|W\|_q
\le
\frac{\mu+cq+\sqrt{(\mu-cq)^2+2vq}}{2}.
$$
\end{lemma}
\begin{proof}
Write the right-hand side as $B$. Then $B>cq$ (since $v>0$) and
$(B-\mu)(B-cq)=vq/2$. Taking $\lambda=q/B$ in the MGF bound gives
$$
\log\E e^{\lambda W}
\le\frac{q\mu}{B}+\frac{vq^2}{2B(B-cq)}=q.
$$
Since $x^q\le(q/(e\lambda))^q e^{\lambda x}$ for $x\ge0$, taking expectations and $q$-th roots gives $\|W\|_q\le B$.
\end{proof}

\begin{remark}
The lemma lets us \emph{center the concentration} at the mean. 
Since $W\ge0$, $\|W\|_q\le\mu+\frac{vq}{2(\mu-cq)}$ for $\mu>cq$, and the extra term is of order $vq/\mu$, constant in our application ($q=d$, $v\asymp\mu/d$). 
Standard moment inequalities do not give this directly: Minkowski's inequality $\|W\|_q\le\mu+\|W-\mu\|_q$ with $\|W-\mu\|_q\lesssim\sqrt{vq}+cq$ leaves $\sqrt\mu$, which forces $\mu\gtrsim\varepsilon^{-2}$ and turn our $d/\varepsilon$ term into $d/\varepsilon^2$.
\end{remark}

\paragraph{Sensitivity of collision counts.}
Throughout this section $d\ge2$ is fixed, and $n\ge d$, since for $n<d$ we have $C=m=0$ and \Cref{thm:concentration} is trivial. We write $C=\widehat C_d$ for the raw collision count and put $m=(\E C)^{1/d}$, so that $\hat H_d-H_d(p)=-\frac1{d-1}\log(C/m^d)$.
It suffices to show, with the relative accuracy $\varepsilon/4$ for the root,
\begin{align}\label{eq:norm_estimation}
|C^{1/d}-m|\le \tfrac\varepsilon4\,m,
\end{align}
as verified at the end of this section.
We will prove the concentration for the \emph{smoothed} quantity
\begin{align}
F = (C+4^d)^{1/d}.
\end{align}
The smoothing changes the root by at most $4$; below we absorb this deterministic error into the allowed relative error.
We now compute the \emph{deletion sensitivity} with respect to the $i$-th sample:
$$
\Delta_i = C - C^{(-i)} = \binom{N_{X_i}-1}{d-1},
$$
because deleting an observation from the cell of $X_i$ with frequency $q\ge d$ removes
$\Delta_i=\binom{q-1}{d-1}$ collisions, leaving $\binom{q}{d}-\binom{q-1}{d-1}=\binom{q-1}{d}$; otherwise it removes zero.

The next step is to handle two regimes: if $N_x < 2 d$ (\emph{low count}), then 
$\Delta_i \leq 4^d/2$. For \emph{high-count} when $N_x \geq 2 d$
by the inequality $\binom{N-1}{d-1}\leq \frac{1}{2}\binom{N}{d}$ valid for $N\geq 2d$
we have $\Delta_i \leq C/2$. Thus, 
$$
\Delta_i \leq \frac{ C + 4^d}{2}.
$$
We now bound the \emph{local sensitivity} of $C$. Put $N=N_{X_i}$.
Using $N\Delta_i=d\binom Nd$ and $\Delta_i\le N^{d-1}/(d-1)!$, we get
$$
\Delta_i^d\le\frac{(N\Delta_i)^{d-1}}{(d-1)!}
=\frac{d^{d-1}}{(d-1)!}\binom Nd^{d-1}
\le e^d\binom Nd^{d-1},
$$
where $d!\ge(d/e)^d$. Therefore
$$
\Delta_i\le e\binom Nd^{(d-1)/d}
\le eC^{(d-1)/d}\le eF^{d-1}.
$$
We obtain that the \emph{total sensitivity} equals exactly
$$
\sum_{i}^{n}\Delta_i = \sum_{x:N_x>0} N_x \binom{N_x-1}{d-1} = d C,
$$
because every collision will be counted $d$ times towards removals.

Define $C^{(-i)}$ by deleting observation $i$, and $F^{(-i)}=(C^{(-i)}+4^d)^{1/d}$. Now transfer these bounds to $F$. By the concavity of $d$-th root,
$$
F-F^{(-i)}  = (C + 4^d)^{1/d} - (C -\Delta_i + 4^d)^{1/d} \leq
\frac{\Delta_i }{d (C - \Delta_i + 4^d)^{\frac{d-1}{d}}} \leq \frac{2 \Delta_i}{d F^{d-1}}
$$
and thus we can bound total sensitivity of $F$
$$
0\leq \sum_i (F - F^{(-i)}) \leq  \frac{2 }{d F^{d-1}}\cdot d C  \leq 2 F.
$$
The local sensitivity of $F$ is bounded by:
$$
F-F^{(-i)} \leq \frac{2 \Delta_i}{d F^{d-1}} \leq 2e / d
$$

\paragraph{Concentration bounds.}
Set \(Z=\frac{d}{2e}F\) and \(Z^{(-i)}=\frac{d}{2e}F^{(-i)}\). The preceding bounds give
$$
0\le Z-Z^{(-i)}\le1,
\qquad
\sum_i(Z-Z^{(-i)})\le2Z.
$$
Since $0\le Z-Z^{(-i)}\le1$, also $\sum_i(Z-Z^{(-i)})^2\le\sum_i(Z-Z^{(-i)})\le 2Z$, so $Z$ is $(2,0)$-self-bounding. By the upper-MGF and lower-tail bounds in \cite[Theorem~1]{BLM09}, writing $\mu=\E F$, we have
$$
\log\E e^{s(F-\mu)}
\le\frac{(2e\mu/d)s^2}{1-2es/d},
\qquad 0<s<d/(2e).
$$
In particular, $F$ is upper sub-gamma with variance proxy $4e\mu/d$ and scale $2e/d$; its lower tail is sub-Gaussian with the same variance proxy. Consequently, for every $t>0$,
$$
\Pr\!\left(|F-\mu|>
\sqrt{\frac{8e\mu t}{d}}+\frac{2et}{d}\right)\le2e^{-t}.
$$
We now transfer these guarantees from the smoothed objective $F$ to the actual collision statistic $C$.
To this end, we apply \Cref{thm:subg_1side_moment} with
$$
W=F=(C+4^d)^{1/d},\qquad
q=d,\qquad
v=\frac{4e\mu}{d},\qquad c=\frac{2e}{d},
$$
where \(\mu=\mathbb EF\). We obtain
$$
(\mathbb EF^d)^{1/d}
\le
\frac{\mu+2e+\sqrt{(\mu-2e)^2+8e\mu}}2
=\mu+2e.
$$
By this and Jensen's inequality 
$$
m\le(\mathbb EF^d)^{1/d}\le\mu+2e,
\qquad
\mu\le(m^d+4^d)^{1/d}\le m+4.
$$
Thus
$$
m-2e\le\mu\le m+4.
$$

Combining this with sub-gamma concentration and
\(F-4\le C^{1/d}\le F\), for every \(t>0\),
$$
\Pr\!\left(
|C^{1/d}-m|
>
4+2e+\sqrt{\frac{8e(m+4)t}{d}}+\frac{2et}{d}
\right)\le2e^{-t}.
$$
This proves \Cref{thm:concentration}.

\paragraph{Sufficient sample size.}
Let $t=\log(2/\delta)$. By the transfer bound, with probability at least $1-\delta$,
$$
|C^{1/d}-m|\ \lesssim\ 1+\sqrt{\frac{m t}{d}}+\frac td .
$$
Since $\binom nd\ge(n/d)^d$, we have $m\ge\frac nd\|p\|_d$. The right-hand side is at most $\varepsilon m/4$ as soon as
$$
m\ \gtrsim\ \frac1\varepsilon+\frac{t}{d\varepsilon^2},
\qquad\text{i.e.}\qquad
n\ \gtrsim\ \|p\|_d^{-1}\Big(\frac d\varepsilon+\frac{\log(1/\delta)}{\varepsilon^2}\Big),
$$
and $\|p\|_d^{-1}\le K$ because $P_d\ge k^{1-d}$. Then $|C^{1/d}/m-1|\le\varepsilon/4$, hence
$$
|\hat H_d-H_d|=\frac d{d-1}\Big|\log\frac{C^{1/d}}{m}\Big|\le\varepsilon .
$$
This proves \Cref{thm:estimator}.

\begin{lemma}[Entropy order cap]\label{lem:cap}
For a distribution on at most $k$ symbols and integers $2\le d'\le d$,
$$
0\le H_{d'}(p)-H_d(p)\le\frac{\log k}{d'-1}.
$$
\end{lemma}
\begin{proof}
Monotonicity of R\'enyi entropy gives $H_\infty\le H_d\le H_{d'}$.
Also $P_{d'}\ge p_{\max}^{d'}$ implies $H_{d'}\le d'H_\infty/(d'-1)$.
Thus $H_{d'}-H_d\le H_\infty/(d'-1)\le\log k/(d'-1)$.
\end{proof}

\begin{proof}[Proof of \Cref{cor:capped}]
If $d'=d$, there is no order bias. Otherwise $d'=\lceil1+2\log k/\varepsilon\rceil$ and \Cref{lem:cap} gives
$0\le H_{d'}-H_d\le\log k/(d'-1)\le\varepsilon/2$.
Apply \Cref{thm:estimator} at order $d'$ with accuracy $\varepsilon/2$, and use
$\|p\|_{d'}^{-1}\le k^{1-1/d'}\le K$.
If the cap does not bind, $d'\varepsilon=d\varepsilon\le2\varepsilon+2\log k\lesssim\log K$, so
$d'\varepsilon\lesssim\min\{d\varepsilon,\log K\}$.
If it binds, $d\varepsilon>2\log k$ and $d'\varepsilon\le2\varepsilon+2\log k\lesssim\log K$, giving the same conclusion.
Here $\varepsilon\le1$, $k\ge2$, and $\log k\le2\log K$ absorb the rounding terms.
The stated sample bound follows.
\end{proof}

\subsection{Learning Spiked Plateaus}\label{sec:lower}

The following calculations prove \Cref{thm:plateau_testing} for the family illustrated in~\Cref{fig:hard-family}.

\paragraph{Testing difficulty.}
A test with small error requires separation between the two sample laws.
For distributions $Q\ll P$ we use $\chi^2(Q\Vert P)=\E_P(\tfrac{dQ}{dP})^2-1$.

\begin{lemma}[$\chi^2$ testing bound]\label{lemma:testing}
Let $P,Q$ be distributions and $T$ an event with $P(T^c)\le\delta$ and $Q(T^c)\ge1-\delta$. Then
$$
1+\chi^2(Q\Vert P)\ \ge\ \frac{(1-\delta)^2}{\delta}.
$$
\end{lemma}
\begin{proof}
By Cauchy--Schwarz, $1-\delta\le Q(T^c)=\mathbf E_P\bigl[\tfrac{dQ}{dP}\mathbf 1_{T^c}\bigr]\le\sqrt{(1+\chi^2(Q\Vert P))\,P(T^c)}$.
\end{proof}

\paragraph{The hard family.}
Choose $S\subset[k]$ with $|S|=r<k$, put $a=1/K$, and assume
$ra\le1/2$ and $a\eta\le1-ra$.
The null has a plateau on $S$ and a uniform floor on its complement:
$$
p_0(x)=
\begin{cases}
a,&x\in S,\\
\dfrac{1-ra}{k-r},&x\notin S.
\end{cases}
$$
For $j\in S$, raise atom $j$ by $a\eta$ and remove that mass uniformly from the floor:
$$
p_j(x)=
\begin{cases}
a(1+\eta),&x=j,\\
a,&x\in S\setminus\{j\},\\
\dfrac{1-ra-a\eta}{k-r},&x\notin S.
\end{cases}
$$

\paragraph{Intuition.}
With $n$ samples, the count $N_j$ of an atom with mass $1/K$ fluctuates by about $\sqrt{n/K}$ around its mean $n/K$; 
a spike at $j$ with relative increase $\eta$ moves the mean by $\eta n/K$; this change is above the fluctuation when $n\gtrsim K\eta^{-2}$, even if the spike location is known.
Error probability $\delta$ requires the shift to exceed $\sqrt{\log(1/\delta)}$ standard deviations, and the union bound over the $r$ candidate locations raises this to $\sqrt{\log r+\log(1/\delta)}$. 
This gives, for $\eta\le1$
$$
n\asymp
\underbrace{K\eta^{-2}}_{\text{relative-change detection}}
\left(
\underbrace{\log(1/\delta)}_{\text{confidence}}
+
\underbrace{\log r}_{\text{search cost}}
\right).
$$

\paragraph{The test.}

The single-sample likelihood ratio is
$$
\ell_j(x)=\frac{p_j(x)}{p_0(x)}=
\begin{cases}
1+\eta,&x=j,\\
1,&x\in S\setminus\{j\},\\
1-\dfrac{a\eta}{1-ra},&x\notin S.
\end{cases}
$$
For a \emph{fixed spike}, only the boosted atom mass and the floor (outside) contribute:
$$
\chi^2(p_j\Vert p_0)
=\E_{p_0}(\ell_j-1)^2
=a\eta^2+\frac{a^2\eta^2}{1-ra}
\le2a\eta^2,
$$
where we utilised $ra\le1/2$ and $a\le1/2$. By independence of the observations:
\begin{align}\label{eq:bound_fixed_spike}
1+\chi^2(p_j^{\otimes n}\Vert p_0^{\otimes n})
=(\E_{p_0}\ell_j^2)^n
\le e^{2na\eta^2}.
\end{align}

For a \emph{random spike},  $j$ is first sampled uniformly from $S$, and then all $n$ observations are drawn from $p_j$;
thus $Q=r^{-1}\sum_jp_j^{\otimes n}$ is a mixture of product measures.
Writing $L_j(X^n)=\prod_{t=1}^n\ell_j(X_t)$, expansion of the squared mixture likelihood and independence give
\begin{align*}
1+\chi^2(Q\Vert p_0^{\otimes n})
&=\E_{p_0^{\otimes n}}\left(\frac1r\sum_jL_j\right)^2\\
&=\frac1{r^2}\sum_{i,j}\E_{p_0^{\otimes n}}[L_iL_j]
=\frac1{r^2}\sum_{i,j}(\E_{p_0}[\ell_i\ell_j])^n.
\end{align*}
Since $\E_{p_0}\ell_j=1$, write $\E_{p_0}[\ell_i\ell_j]=1+\rho_{ij}$, where
$$
\rho_{ij}=\E_{p_0}[(\ell_i-1)(\ell_j-1)]
=\mathbf1_{\{i=j\}}a\eta^2+\frac{a^2\eta^2}{1-ra}.
$$
The first term contributes when the spikes coincide; the second is the mass deficit outside $S$.
Put $A=a\eta^2$ and $B=a^2\eta^2/(1-ra)\le2a^2\eta^2$.
Separating the $r$ diagonal terms from the $r(r-1)$ off-diagonal terms yields, for every $n$,
\begin{align}\label{eq:bound_mixed_spike}
1+\chi^2(Q\Vert p_0^{\otimes n})
&=\frac1r(1+A+B)^n+\left(1-\frac1r\right)(1+B)^n\notag\\
&\le e^{nB}\left[1+\frac{e^{nA}-1}{r}\right]
\le e^{2na^2\eta^2}\left[1+\frac{e^{na\eta^2}-1}{r}\right].
\end{align}

\paragraph{Distinguishing alternative from null.}
Assume $\delta\le1/8$ and put $z=na\eta^2=n\eta^2/K$.
For a test between $P=p_0^{\otimes n}$ and $Q$ with both errors at most $\delta$, the testing lemma requires
$$
1+\chi^2\ge\frac{(1-\delta)^2}{\delta}\ge\frac1{2\delta}.
$$
For the confidence cost, the bracket in \eqref{eq:bound_mixed_spike} is at most $e^z$, so
$$
1+\chi^2(Q\Vert P)\le e^{(1+2/K)z}\le e^{2z},
\qquad
z\ge\frac12\log\frac1{2\delta}\ge\frac13\log\frac1\delta.
$$
For the \emph{random spike}, suppose instead that $z\le\frac18\log r$.
Since $r\le K/2$ and $\log r\le r$, \eqref{eq:bound_mixed_spike} gives
$$
1+\chi^2(Q\Vert p_0^{\otimes n})
\le e^{2z/K}\left[1+\frac{e^z-1}{r}\right]
\le 2e^{1/8}<4\le\frac1{2\delta},
$$
a contradiction. Hence $z>\frac18\log r$.
Combining the confidence and search bounds gives the necessary sample size
\begin{equation}\label{eq:plateau_sample_lower}
n\gtrsim\frac{K}{\eta^2}
\left(\log r+\log\frac1\delta\right).
\end{equation}
This proves the lower bound in \Cref{thm:plateau_testing}, with no restriction on $\log(1/\delta)$.

For the converse, let $0<\eta\le1$, let $N_j$ be the sample count of $j$, and $\phi=1$ (reject) if $\max_{j\in S}N_j\ge na(1+\eta/2)$. Under $P$, each $N_j\sim\mathrm{Bin}(n,a)$, so the multiplicative Chernoff bound and a union bound give $P(\phi=1)\le r\,e^{-na\eta^2/16}$. Under $p_j$, $N_j\sim\mathrm{Bin}(n,a(1+\eta))$, so $p_j^{\otimes n}(\phi=0)\le e^{-na\eta^2/16}$, and the same bound holds for $Q(\phi=0)$ by averaging over $j$. 
Both errors are at most $\delta$ if $n\ge16K\eta^{-2}\log(r/\delta)$, which matches \eqref{eq:plateau_sample_lower}. This completes the proof of \Cref{thm:plateau_testing}.

\subsection{R\'enyi Entropy Lower Bounds}\label{sec:entropy_lower}

We apply the hidden-spike testing bound to entropy estimation, using the same plateau family.

\begin{corollary}[R\'enyi entropy lower bound]\label{thm:plateau_lower}
Let $d\ge2$ be an integer, $K=k^{1-1/d}\ge4$, $0<\varepsilon\le1/64$, and $0<\delta\le1/8$.
Every estimator that is $\varepsilon$-accurate with confidence $1-\delta$ over the $k$-symbol simplex requires
$$
n\gtrsim\frac{K}{\varepsilon^2}
\left(\min\{d\varepsilon,\log K\}+\log\frac1\delta\right).
$$
\end{corollary}

\paragraph{From estimation to testing.}
\begin{lemma}
If $\hat H$ is $\varepsilon$-accurate with confidence $1-\delta$ and $H_d(p_0)-H_d(p_j)>2\varepsilon$ for all $j\in S$, then \Cref{lemma:testing} applies to $P=p_0^{\otimes n}$, $Q=\frac1r\sum_{j\in S}p_j^{\otimes n}$, with $T=\{\hat H\ge H_d(p_0)-\varepsilon\}$ (accuracy gives $P(T^c)\le\delta$ and $p_j^{\otimes n}(T)\le\delta$ for every $j$, hence $Q(T)\le\delta$ by averaging).
\end{lemma}

\paragraph{Parameters.}
Let $0<\varepsilon\le1/64$, $0<\delta\le1/8$, and $K\ge4$. Choose
$$
\gamma=\min\{d\varepsilon,\log(K/4)\},\qquad
r=\lceil e^\gamma\rceil,\qquad
a=\frac1K,\qquad \eta=e^{8\varepsilon}-1.
$$
Then $8\varepsilon\le\eta\le10\varepsilon$, $ra\le1/4+1/K\le1/2$, and $a\eta<1/2\le1-ra$, so the family is valid.
Also $r\le e^{d\varepsilon}+1$.

\paragraph{Entropy.}
We show $H_d(p_0)-H_d(p_j)>2\varepsilon$ for every $j\in S$ (a gap $\gtrsim\varepsilon$ would not suffice: the test needs more than $2\varepsilon$).
Put $h=a\eta$ and $b=\frac{1-ra}{k-r}$. Since $b\le\frac1k$ and $a=\frac1K$, we have $(b/a)^{d-1}\le\frac1K$.
By region,
$$
P_d(p_j)-P_d(p_0)=
\underbrace{(a+h)^d-a^d\vphantom{(a+h)^d-a^d0(k-r)\Bigl[b^d-\bigl(b-\tfrac{h}{k-r}\bigr)^d\Bigr]}}_{x=j}
+\underbrace{0\vphantom{(a+h)^d-a^d0(k-r)\Bigl[b^d-\bigl(b-\tfrac{h}{k-r}\bigr)^d\Bigr]}}_{x\in S\setminus\{j\}}
-\underbrace{(k-r)\Bigl[b^d-\bigl(b-\tfrac{h}{k-r}\bigr)^d\Bigr]\vphantom{(a+h)^d-a^d0(k-r)\Bigl[b^d-\bigl(b-\tfrac{h}{k-r}\bigr)^d\Bigr]}}_{x\notin S}.
$$
By the mean value theorem, the last term is at most $d\,b^{d-1}h\le\bigl(\tfrac ba\bigr)^{d-1}\bigl[(a+h)^d-a^d\bigr]\le\frac1K\bigl[(a+h)^d-a^d\bigr]$, so
$$
P_d(p_j)-P_d(p_0)\ \ge\ \Bigl(1-\tfrac1K\Bigr)a^d\bigl[(1+\eta)^d-1\bigr].
$$
Since $P_d(p_0)\le(r+1)a^d$, with $\eta=e^{8\varepsilon}-1$, $u=d\varepsilon$, $r+1\le3e^{u}$ and $K\ge4$,
$$
\frac{P_d(p_j)}{P_d(p_0)}\ \ge\ 1+\frac{e^{8u}-1}{4e^{u}}\ \ge\ e^{2u},
$$
because $\frac{e^{8u}-1}{e^{2u}-1}=1+e^{2u}+e^{4u}+e^{6u}\ge4e^{3u}\ge4e^{u}$. Hence
$$
H_d(p_0)-H_d(p_j)=\frac1{d-1}\log\frac{P_d(p_j)}{P_d(p_0)}\ \ge\ \frac{2d\varepsilon}{d-1}\ >\ 2\varepsilon .
$$

\paragraph{Lower bound.}
\begin{proof}[Proof of \Cref{thm:plateau_lower}]
The entropy argument gives $H_d(p_0)-H_d(p_j)>2\varepsilon$ for every $j$.
Thus accuracy yields a test between the null and the mixture with both errors at most $\delta$, and \Cref{thm:plateau_testing} applies.
Substituting $\eta\le10\varepsilon$ gives
$$
n\gtrsim\frac{K}{\varepsilon^2}\left(\log r+\log\frac1\delta\right).
$$
Finally, $\log r\ge\gamma$ and
$\min\{d\varepsilon,\log K\}\le\gamma+\log4$.
The additive $\log4$ is absorbed by $\log(1/\delta)\ge\log8$, giving the stated bound.
\end{proof}

\begin{proof}[Proof of \Cref{thm:main}]
\Cref{cor:capped} gives the upper bound with the stated capped estimator.
For $k\ge16$, $K\ge\sqrt k\ge4$, so \Cref{thm:plateau_lower} gives the matching lower bound throughout the stated parameter range.
\end{proof}

\section{Discussion}

% NOTE: one-sentence summary of the result, so the two ingredients below read as its two parts.
We settled the sample complexity of R\'enyi entropy estimation for integer orders with \emph{direct} birthday estimators, using two new ingredients.

% NOTE: ingredient 1, named, with the theorem that delivers it.
The first is a direct concentration bound for the collision count (\Cref{thm:concentration}).
% NOTE: what the standard approach costs: the confidence term multiplies the whole base sample size.
The usual routes to high confidence, median aggregation over blocks and blocking arguments, multiply the whole base sample size by $\log(1/\delta)$; blocking also loses the birthday scale.
% NOTE: what we get instead: confidence is additive, which is the shape of the final rate.
Our bound has a Bernstein-type tail, so that confidence impacts the sample size additively in the entropy guarantee (\Cref{thm:estimator}).
% NOTE: why the method is cheap and portable: it needs only deletion sensitivity.
It uses only the deletion sensitivity of the count, and may extend to other statistics built from coincidences.

% NOTE: ingredient 2, named: the three-level family.
The second is on testing spiked plateaus: a uniform floor, a plateau, and one hidden spike.
% NOTE: what we prove about it: testing is tight, and the two knobs (r, eta) recover all three regimes, i.e. the lower bound.
This family complexity is determined up to constants (\Cref{thm:plateau_testing}), and tuning the plateau size and the boost gives matching lower bounds in all three regimes of the entropy rate.
% NOTE: positioning against earlier constructions, and the independent-interest claim.
It differs from analogous constructions in prior work, and may be of independent interest wherever a relative change of a heavy atom must be detected among many candidates.

% NOTE: future work, not a limitation of this paper: high confidence and adaptivity together.
A natural direction for future work is high-confidence \emph{and} adaptive R\'enyi entropy estimation: estimators that attain the instance-dependent rate in terms of $\|p\|_d^{-1}$ with failure probability $\delta$, without prior knowledge of the distribution, together with matching lower bounds.

\section*{AI use disclosure}
An AI assistant (Claude, Anthropic) was used to support literature search and help polish an early version of the text.
The author takes responsibility for all content.

\bibliographystyle{plain}
\bibliography{citations}

\end{document}